\documentstyle[]{mn}

\newlength{\colwidthf}
\newlength{\hcolwidthf}
\newlength{\colwidth}
\newlength{\hcolwidth}
\input ./psfig.sty
\newif\ifpsfiles\psfilestrue
\psfilestrue
\def\getfig#1#2{\ifpsfiles\psfig{figure=#1,width=\hsize}\else\vskip#2\fi}
\def\get2fig#1#2{\ifpsfiles\psfig{figure=#1,width=#2}\else\vskip#2\fi}

\def\phibar{\varphi_{\rm bar}}

\def\kpc{\,{\rm kpc}}

\def\spose#1{\hbox to 0pt{#1\hss}}
\def\lta{\mathrel{\spose{\lower 3pt\hbox{$\mathchar"218$}}
     \raise 2.0pt\hbox{$\mathchar"13C$}}}
\def\gta{\mathrel{\spose{\lower 3pt\hbox{$\mathchar"218$}}
     \raise 2.0pt\hbox{$\mathchar"13E$}}}
\def\pc{{\rm\,pc}}
\def\kpc{{\rm\,kpc}}

\title[Non-parametric versus parametric modelling]
{Parametric versus non-parametric modelling? Statistical evidence
based on $P$-value curves.}

\author[N. Bissantz, A. Munk and A. Scholz]{Nicolai Bissantz$^1$, Axel 
Munk$^1$ and Achim Scholz$^{1,2}$\\
$^1$Institut f\"ur Mathematische Stochastik, Universit\"at G\"ottingen, Lotzestr.13,
D-37083 G\"ottingen, Germany\\
$^2$Citibank Privatkunden AG, D\"usseldorf, Germany}
\def\3{\ss}

\newcommand{\bea}{\begin{eqnarray*}}
\newcommand{\eea}{\end{eqnarray*}}
\newcommand{\be}{\begin{eqnarray}}
\newcommand{\ee}{\end{eqnarray}}

\begin{document}
\maketitle


\begin{abstract}
In astrophysical (inverse) regression problems
it is an important task to decide whether a given
parametric model describes the observational data sufficiently
well or whether a non-parametric modelling becomes necessary.
However, in contrast to common practice this cannot be decided
by solely comparing the quality of fit due to possible 
over-fitting by the non-parametric method. Therefore, 
in this paper we present a resampling algorithm which allows
to decide whether deviations between a parametric 
and a non-parametric model 
are systematic or due to noise. 
The algorithm is based on a statistical comparison of the
corresponding residuals, under the assumption of the parametric
model as well as under violation of this assumption. This
yields a graphical tool for a robust decision making of parametric
versus non-parametric modelling.  

Moreover, our approach can be used for the selection of the
most proper model among several competitors (model selection).
The methods are illustrated by the problem of recovering the
luminosity density in the Milky Way [MW] from near-infrared [NIR] surface
brightness data of the {\sl DIRBE} experiment on board of the 
{\sl COBE} satellite. Among the parametric models investigated
one with 4-armed spiral structure performs
best. In this model the Sagittarius-Carina arm and its 
counter-arm are significantly weaker than the other pair of arms.
Furthermore, we find statistical evidence for an improvement
over a range of parametric models with different spiral structure
morphologies by a non-parametric model of Bissantz \& Gerhard 
\cite{bi02}. 
\end{abstract}


\section{Introduction}
One of the basic problems in astrophysical research consists
in the proper choice of a model to describe an observational
array $\omega_{\rm obs}(t_i), i=1,\ldots,N$ of $N$ measurements. Here 
$t=(t_1,\ldots,t_N)$ denotes a quantity which affects $\omega=
(\omega(t_1),\ldots,\omega(t_N))$ 
in a systematic, but blurred way, $\omega(t_i)$, the 
regression function to be reconstructed from the data. 
For example, $\omega_{\rm obs}(t_i)$ could be measurements of the 
surface brightness at sky position $t_i=(l_i,b_i)$. In noisy
inverse models $\omega$ itself is not the quantity of primary interest,
rather a function $\rho$ has to be recovered, where the relation
between $\omega$ and $\rho$ is given by a (linear) operator $K$ (matrix), 
viz.
\bea
\omega(t_i)=\left(K\rho\right)(t_i), \qquad i=1,\ldots,N.
\eea
Often $K$ is given by a $N\times N$ matrix, which will in
general be numerically difficult to invert. This will also be 
the case in this paper where we are concerned with 
the recovery of  the spatial 
(three-dimensional) luminosity density of a galaxy from 
blurred observations of its surface brightness and
reverberation mapping of gas in AGNs. For more examples 
of inverse problems in astrophysics see e.g. Lucy \cite{lucy1994}.
Due to the noisy measurements it is tempting to assume that
$\omega_{\rm obs}(t_i)=\omega(t_{i})+\varepsilon_{i},$ 
where the $\varepsilon_{i}$ denote some random
noise and $\omega(t_{i})$ the expected value of $\omega_{\rm obs}(t_i)$, i.e. 
$E[\omega_{\rm obs}(t_i)]=\omega(t_i).$
In particular we allow for different error distributions of the 
$\varepsilon_{i}$, which entails inhomogeneous variance patterns
(cf. Hocking, 1996, for a thorough discussion of models with 
inhomogeneous variances),
viz. $V[\varepsilon_{i}]=\sigma_{i}^2$, as will be the case in our example
of de-projecting the de-reddened {\sl COBE/DIRBE} L-band surface brightness 
map of Spergel et al. \cite{sp95}. Cf. Bissantz \& Munk, \cite{bi01} [BM1] for the
noise properties of this data. 

In this paper we are concerned with a new method to compare
several competing models for the regression function $\rho$ and
to select the most appropriate one. 
These models may be of a certain parametric form (parametric model)
\bea
U=\left\{\rho_{\vartheta}\right\}_{\vartheta\in\Theta}, \qquad 
\Theta\subseteq  I\!\!R^d
\eea
or non-parametric, i.e. only qualitative smoothness or geometric
assumptions such as symmetry or differentiability, (cf. Wand \& Jones
(1995) or Wahba (1990) for a good introduction to non-parametric 
modelling by kernel or spline methods, respectively) are made a-priori.
Implicitly, any algorithm, to reconstruct $\rho$ from $\omega$, 
relies on those assumptions or combinations thereof. Statistical methods
for model selection are broadly used in astrophysics (cf. Feigelson
\& Babu, 2002),
a good statistical introduction into this area is Burnham \& Anderson (1998)
or Eubank (1999) among many others. The case of inhomogeneous inverse models,
as in our application, is not treated explicitely in the literature so far. 

Our approach is based on the statistical comparison of the estimated
parametric residuals 
\bea
\hat\varepsilon_i=\omega_{\rm obs}(t_i) - \left(K\rho_{\hat\vartheta}\right)(t_i), 
\qquad i=1,\ldots,N,
\eea
where $\rho_{\hat\vartheta}$ denotes the best possible fit of the
model class $U$ to the data $\omega_{\rm obs}$ (e.g. obtained by least squares), 
and the estimated non-parametric residuals
\bea
\varepsilon^{\rm (np)}_i=\omega_{\rm obs}(t_i)-\left(K\rho^{\rm (np)}\right) (t_i),
\qquad i=1,\ldots,N,
\eea
where $\rho^{\rm (np)}$ denotes a non-parametric reconstruction
of $\rho$.
This estimator can be obtained by several methods, including 
penalized maximum likelihood if the error distribution of 
$\varepsilon$ is known (e.g. Magorrian et al., \cite{ma98}, Bissantz \&
Gerhard, 2002), the Richardson-Lucy iterative 
method (e.g. Binney, Gerhard \& Spergel, 1997 [BGS]), 
subtractive optimally localized averages (e.g. Pijpers \& Thompson, 1994, 
Pijpers \& Wanders, 1994), and
maximum entropy methods (e.g. Wallington et al., 1994, 1996). 

In our example, which will be discussed in detail in Sect. \ref{basics} and \ref{appl},
a penalized maximum likelihood method is used, where the penalty
terms encourage symmetry, smoothness features and spiral structure
of the recovered density distribution. In this model the error distribution
is assumed to be independent of the data point.

Now the general methodology will be to use the difference of the residuals
$\hat\varepsilon_i$
and $\varepsilon^{\rm (np)}_i$
\bea
\hat\delta_i = \hat\varepsilon_i - \varepsilon^{\rm (np)}_i = \left(K\rho_{\hat\vartheta}-
K\rho^{\rm (np)}\right)(t_i).
\eea
Roughly speaking, a small sum of squares of $\hat\delta_i$
will indicate that the non-parametric and the parametric fit are close, 
which should give evidence for the parametric model to hold. Otherwise, if 
$\frac{1}{N}\sum \hat\delta_i^2$ is large the non-parametric fit outperforms the
parametric one. 
Due to the possibly inhomogeneous variance pattern of the error $\sigma_i^2$,
a valid statistical analysis requires that $\hat\delta_i$ has to be
weighted with the estimated local variability 
$\hat\sigma^2_i=\hat\sigma^2(t_i)$, which results in a {\bf l}ocally {\bf w}eighted
{\bf r}esidual {\bf s}um of {\bf s}quares
\bea
{\rm LWRSS}=\frac{1}{N} \sum\limits_{i=1}^N \left(\frac{
\hat\varepsilon_i - \varepsilon^{\rm (np)}_i}{\hat\sigma_i}\right)^2.
\eea
Observe, that the use of the locally weighted residual sums of squares
leads to a quantity which does not depend on the (unknown) local variability
of the data, i.e. $E\left[(\hat\varepsilon_i-\varepsilon^{\rm (np)}_i)/
\hat\sigma_i \right]^2$ will be a quantity which is independent of 
$\sigma_i^2$ for large $N$.
In fact, we claim that under certain regularity conditions on $K$,
the error distribution of $\varepsilon$, $\sigma$ and $\rho$,
$N\sqrt{h_N}{\rm LWRSS}$ has a normal limit, where $h_N$ is a 
sequence tending to zero as $N\rightarrow \infty$.
Hence the distribution of $N\sqrt{h_N}{\rm LWRSS}$ for large  
numbers of observations tends to be asymptotically normal 
with a rather complicated variance which will depend on $K$, $\sigma$
and the smoothing method used for obtaining $\hat\rho$. For the case
of direct regression (then $K$ is the identity) this was made explicite
by Dette \cite{de99} and Dette \& Munk \cite{de02}, 
where we claim that the proof in the indirect case follows
a similar pattern, and is postponed to a different paper. In order to base
a proper decision whether $\rho_{\hat\vartheta}$ is acceptable or 
$\rho^{\rm (np)}$ should be preferred, the approximate normal distribution
of $N\sqrt{h_N}{\rm LWRSS}$
can now be used. The only problem which remains is to determine the 
variance of $N\sqrt{h_N}{\rm LWRSS}$ 
which will be done in the sequel by a resampling (bootstrap)
algorithm from the data. Various simulation studies have shown that this
method even leads to a better approximation of the true distribution of 
RSS (for finite $N$) than 
the asymptotic $(N\rightarrow\infty)$ normal law. 
This is in accordance with work of Dette, von Lieres und Wilkau \& Sperlich 
\cite{de01}, who investigated various variants of this algorithm in a
different context and came to the same conclusion. 
Our bootstrap algorithm will be presented in Sect. \ref{bootstrap}.

In Sect. \ref{appl} we apply the algorithm to the analysis of a dust-corrected 
near-infrared [NIR] {\sl COBE/DIRBE} L-band map of the Milky Way [MW]
(Spergel et al., 1995).
Bissantz \& Gerhard \cite{bi02} modelled this data non-parametrically
with an implementation of the penalized maximum likelihood method. 
We find that their model improves significantly on various parametric models 
constructed from the parametric model of the same data 
[BGS], supplemented by different spiral structure models
(cf. Sect. 4 for a description). 
By our method it can be concluded that this
improvement in fit is not due to overfitting of the data, but rather to 
systematic departures between data and parametric models, which are in 
particular not flexible enough to capture certain deviations from a
double-exponential disk and smooth spiral arms.

We mention that our method can be used as well to decide between
several classes of concurring parametric models (with possibly different
numbers of parameters). Among the parametric models with different spiral
structure that we have investigated, a four-armed model with the 
Sagittarius-Carina arm (and its counter-arm) significantly weaker than the 
other arms (cf. Drimmel \& Spergel, 2001), outperforms its competitors.

Finally, we mention that the main difference of our method to previous work
of the authors ([BM1], Bissantz \& Munk, 2002 [BM2]) consists in three important
aspects. First, LWRSS adapts to local variability estimated 
from data, which yields an overall measure of goodness of fit weighting the
impact of data due to its local variability. Our numerical analysis shows 
that this yields much more reliable results as before. Second, we do not require
an additional smoothing step for the residual differences as in [BM2]. 
This smoothing step had to be introduced in [BM1] in order
to obtain a distributional limit of the statistic considered there. 
However, due to this additional smoothing the statistics becomes much harder to
interpret. In the present approach this is not necessary anymore, and our
statistics LWRSS measures asymptotically the $L^2$-distance between the true
model $\rho$ and the parametric model $\rho_{\vartheta}$, i.e. 
\bea
M^2=\min_{\vartheta\in\Theta} ||K(\rho-\rho_{\vartheta})||^2
\eea
where $||\cdot||$ refers to the euclidian norm. Furthermore, our 
new method also allows comparison with a non-parametric 
competitor, which serves as an objective alternative.

For those readers, who are mainly interested in the methodological part
of this paper we recommend to skip Sect. 4.

\section{A new statistical method to decide between parametric
and non-parametric modelling}
\label{sec2}
As mentioned in the introduction the underlying idea of our approach
is to base the decision, whether the model $U$ should be considered
as acceptable, on the LWRSS, which has to be computed in the following
steps. We illustrate our algorithm for the case where $t_i$ are located
on a two-dimensional grid as it will be the case in our example in 
Sect. \ref{basics}. 
\begin{enumerate}
\item (Computation of the residual difference). Observe, that 
$\hat\varepsilon_i - \varepsilon^{\rm (np)}_i=\omega_{\hat\vartheta}-
\omega^{\rm (np)}$. 
\item (Computation of a local estimate $\hat\sigma^2$ of the variance 
$\sigma^2$). 
Here, several methods are appropriate. A simple method is based
on Savitzky-Golay filters (Press et al., 1994), 
which are common in astrophysics.
The method requires the computation of the differences
between a non-parametric model and the observed data
at $7$ neighbouring points on a grid:
\bea
\hat\sigma^2(l_i,b_j)&=&\sum\limits_{k=-2}^2 c_k\{\nonumber\\
&&
\sum\limits_{l=-2}^2 c_l \left(\omega^{\rm (np)}(l_{i+l},b_{j+k})-\omega_{\rm obs}
(l_{i+l},b_{j+k})\right)^2
\}\nonumber
\eea
\bea
i=3,\ldots,n-2 \qquad {\rm and} \qquad j=3,\ldots,m-2
\eea
where the $\omega_{\rm obs}(l_i,b_j),\quad ^{i=1,\ldots,n}_{j=1,\ldots,m}$
are the observations, the $c_{k}$ and $c_l$ are weights chosen according to 
a Savitzky-Golay filter (Press et al., 1994, setting $n_L\!=\!n_R\!=
\!M\!=\!2$) and $\omega^{\rm (np)}$ is a non-parametric model 
of the Milky Way.

\begin{figure}
\getfig{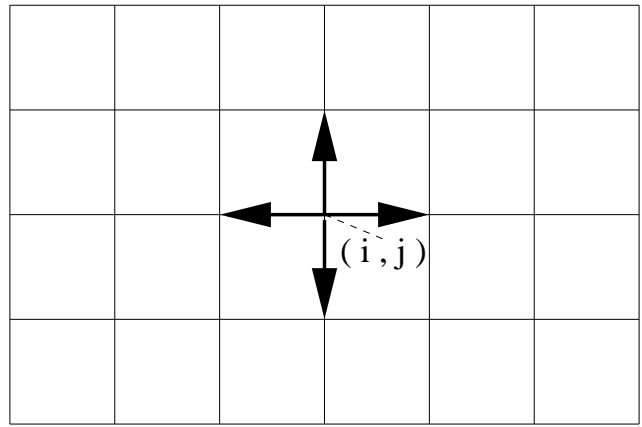}{6cm}
  \caption{Local residuals required for estimating the variance $\sigma^2$.}
\label{figprob}
\end{figure}

Further approaches in the estimation of the local variability are possible with
methods based on the computation of the local differences
of, say $4$, neighbouring points on a grid (cf. Fig. \ref{figprob}): 
\begin{eqnarray}
\hat\sigma_{ij}^2&=&\frac{1}{8}\{
\sum\limits_{k=-1}^1 \left(\omega_{\rm obs}(l_i,b_j)-\omega_{\rm obs}
(l_{i+k},b_j)\right)^2 \nonumber\\
&+& \sum_{k=-1}^1\left(\omega_{\rm obs}(l_i,b_j)-\omega_{\rm obs}
(l_i,b_{j+k}) \right)\},\nonumber
\end{eqnarray}
\bea
i=2,\ldots,n-1 \qquad {\rm and} \qquad j=2,\ldots,m-1.
\eea
Then, in a second step the average over a window of neighbouring local residuals
$\hat\sigma_{ij}^2$ is computed, viz. 
\bea
\hat\sigma^2\equiv \sum\limits_{i,j=-r,-s}^{r,s} \hat\sigma^2_{ij},
\eea
where the window size can be chosen according to prior information on 
constant regions of the expected local variability.

Better but more computer intensive 
methods can be obtained by applying a kernel estimator to the residual
squares $\hat\varepsilon^2_i$, see e.g. Ruppert et al. \cite{rup97}, or biased 
reduced variance estimators (Thompson et al., 1991, 
Munk et al., 2001). 
\item (Compute LWRSS). 
Now, the distribution of LWRSS is required, which is extremely difficult
to compute explicitely. See, e.g. for the case of a direct regression
model, Dette \cite{de99}. Nevertheless, in the following we describe a resampling
algorithm which performs well. Numerical investigations have shown (cf. 
Dette, 1999, and Dette et al., 2001) that a modification of the wild 
bootstrap approximates the true distribution of LWRSS very well and a bootstrap
limit law has been proved in Dette \& Neumeyer (2001) in a similar but simpler
context. Even for 
rather small numbers of observations, $N=100$, say, these authors found in a 
broad range of scenarios on the error distribution and $\rho$ satisfactory
results. This is in accordance with our findings. 
\end{enumerate}

\section{Performing the bootstrap algorithm}
\label{bootstrap}
Our proposed algorithm is a modification of the algorithm presented in 
Bissantz \& Munk \cite{bi01}, and consists of the following steps. \\

{\bf Step 0:} Compute $\omega^{\rm (np)}$.\\
{\bf Step 1:}  ({\sl Generate random data}). 
Generate $N$ random data $\omega_{\rm obs}^{(i)}$ by drawing with 
replacement from the observed data $\omega_{\rm obs}$.\\
{\bf Step 2:} ({\sl Fitting of the random data}).
Determine the ''parametrically best-fitting model''
$\rho_{\hat\vartheta^{(i)}}$ of
the random data $\omega_{\rm obs}^{(i)}$. \\
{\bf Step 3:} ({\sl Compute the target}).
Compute the LWRSS $\hat M^{2}_{(i)}=||\omega_{\hat\vartheta^{(i)}}-
\omega^{\rm (np)}||^2_{\rm LWRSS}$ (see below for the formal definition
of this distance measure).\\
{\bf Step 4:} ({\sl The replication process}).
Repeat steps $1-3$ $B$ times. This yields values
$\hat M^{2}_{(1)},\ldots,\hat M^{2}_{(B)}$. $B$ is a large
number, with $B\approx 500-1000$ usually sufficient. In 
contrast to [BM2] the non-parametric fit $\omega^{\rm (np)}$
is kept fixed, whereas the parametric fit is randomly perturbed
in this algorithm. This leads to a significant reduction of
computing time.

Observe that $\hat M^{2}_{(1)},\ldots, \hat M^{2}_{(B)}$ are 
realisations of a random quantity $X=\hat M^2_{\ast}$, and 
the cumulative distribution function $F^{\ast}_B$ approximates
the distribution function $F$ of $\hat M^2$. 

This algorithm is in accordance with the algorithm of Dette et al.
\cite{de01} who suggested this method in a different context. Interestingly,
these authors came to the same conclusion, that bootstrapping from the
parametric model gives much more reliable results than bootstrapping
from the non-parametric residuals $\varepsilon^{\rm (np)}_i$. This is an 
empirical finding and somehow in contrast to theoretical results.

\section{Near-infrared models of the Milky Way}
\label{basics}
We now apply our proposed method to models of {\sl COBE/DIRBE} NIR data
in the following.
Such models of the distribution of near-infrared luminosity in the 
inner MW are particularly interesting because NIR light traces luminous 
mass well (e.g. Rix \& Zaritzky, 1995). 
{\sl COBE/DIRBE} NIR maps were used to estimate the NIR luminosity distribution 
both parametrically (e.g. Freudenreich, 1998, Dwek et al., 1995, and
Drimmel \& Spergel, 2001),
and non-parametrically ([BGS], Bissantz, Englmaier,
Binney \& Gerhard, 1997, and Bissantz \& Gerhard, 2002).
Only the models of Drimmel \& Spergel \cite{dr01} and Bissantz \& Gerhard \cite{bi02}
contain spiral arms. Bissantz \& Gerhard \cite{bi02} show that 
inclusion of spiral structure is important for the bar/bulge of the model, as  
only then the elongation of the bulge/bar in their models is
large enough to reproduce clump giant star count data of Stanek et al. 
(1994, 1997). Such star count data contain information about
the distances to the surveyed stars, complementary to the all-sky coverage
of the {\sl COBE/DIRBE} NIR map; thus the star count data provides an
important {\sl a posteriori} test of the model(s).

However, despite of its importance, the morphology of the stellar spiral arms
of the Milky Way (i.e. in the  distribution of luminous mass) is not well known. 
In particular it is unclear whether it is predominantly 2- or
4-armed. Ortiz \& L\'epine \cite{or93} used a 4-armed model with logarithmic 
spiral structure in their model of MW NIR starcounts. However, there is a 
tangent point at $\approx 49\deg$ in the (very probably 4-armed) distribution 
of gas and dust (Englmaier \& Gerhard, 1999),  
which seems to be missing in the {\sl COBE/DIRBE} K-band map of the MW (Drimmel,
2000). This indicates 2-armed rather than 4-armed structure. 
On the other hand this tangent is also weak in CO, possibly due to the 
geometry of the line-of-sight through this arm (Dame, private communication).
In Drimmel \& Spergel \cite{dr01} the analysis of the {\sl COBE/DIRBE} $240\mu m$
(tracing dust) and NIR (tracing stellar light) data was extended, and a combined 
dust and stellar disk model produced. The authors found a best model for the
stellar disk which is 4-armed. However, the Sagittarius-Carina arm in their model
is weaker by a factor of $0.4$ in arm-interarm density contrast than the other
arms. Bissantz \& Gerhard \cite{bi02} found in their non-parametric reference model 
with spiral arms that inclusion of spiral structure significantly 
improves their model, however they were not able to decide whether
2 or 4-armed structure is preferable. 

\begin{figure*}
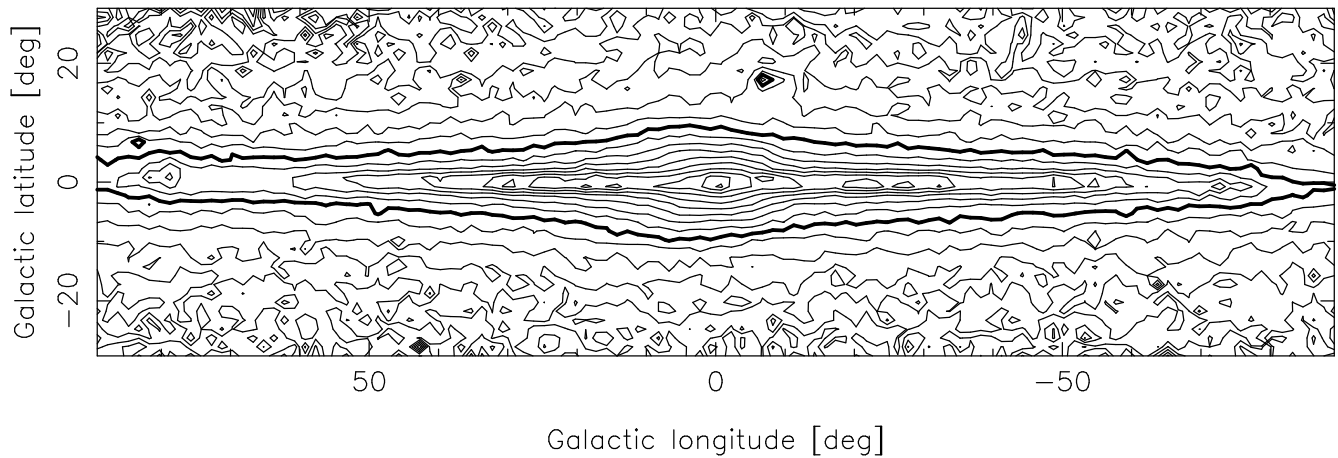

\getfig{cobedaten.ps}{6cm}
\caption{The dust-corrected {\sl COBE/DIRBE} L-band map of
Spergel et al. (1995). 
Contour spacing is 
$0.5 {\rm mag}^2$
and the bold contour is at $0 {\rm mag}^2$. 
Note that there is an arbitrary offset to this scale, however
it is the same as used in Fig. \ref{rmsplot} to allow for 
a comparison of models and data.}
\label{cobedaten}
\end{figure*}

Here we aim to better understand stellar spiral structure in the MW by further 
analysis of a dust-corrected {\sl COBE/DIRBE} L-band map 
(Spergel et al. 1995), shown in Fig. \ref{cobedaten}. 
For our analysis we compare 
four classes of parametric models based on this 
data, which differ with regard to the spiral structure. We include 
models with 2-arms, with 4-arms, without arms and an intermediate model 
consisting of a strong and a weak pair of arms (the latter model as suggested 
by Drimmel \& Spergel, 2001).

We apply our proposed statistical method to perform the comparison of the 
parametric models by using the non-parametric reference model of Bissantz \& 
Gerhard \cite{bi02} as an objective standard of measurement. 
To this end the $P$-value curves of [BM2] are modified for the LWRSS-statistic.
Additionally, with our methodology we find that the non-parametric model better
reproduces the informative part of the L-band map than the best parametric model,
particularly in the disk near spiral arm tangent point features (cf. Fig.
\ref{rmsplot}).
This is achieved by using the distribution of LWRSS which is found by the bootstrap
algorithm in Sect. \ref{bootstrap}. In fact this shows that there is rare evidence
that the difference between the parametric and non-parametric fit is solely 
due to noise, rather than due to a systematic departure of the true
$\rho$ and the best parametric model as described in the last paragraph. 

First we introduce the parametric models in Sect. \ref{para} and the 
non-parametric reference model of Bissantz \& Gerhard \cite{bi02} in 
Sect. \ref{nonpara}. 



\subsection{The parametric models}
\label{para}
The parametric models are defined on a Galacto-centric
Cartesian coordinate system
with axes $x,y,z$, where $x$ is along the major axis, and $y$ along the minor axis 
of the bulge/bar, both in the main plane of the MW. The position of the Sun
in this coordinate system is $z_{\odot}\!=\!14\pc$ above the main plane of the disk 
and $R_{\odot}\!=\! 8\kpc$ from the Galactic centre, 
and the angle between the major axis of the bar and the line-of-sight from the Sun 
to the Galactic Centre is $\phi_{\rm bar}\!=\!20\deg$ [BGS]. 

The parametric models analysed in this paper are similar, except for their spiral 
structure. The other model constituents are a double-exponential disk and a truncated 
power-law bulge (cf. [BGS]). Calling the disk density $\rho_d$ and the bulge density 
$\rho_b$ we define the model density $\rho$ as:
\be
\label{eqpara}
\rho(\vec{x}) = \rho_{d}(\vec{x}) + \rho_{b}(\vec{x}),
\ee
where
\be
\rho_{d}\equiv \rho_d^0 \cdot R_d \cdot e^{-R/R_d} \cdot
\left(\frac{e^{-|z|/z_0}}{z_0}+\alpha \frac{e^{-|z|/z_1}}{z_1}\right),
\nonumber
\ee
\be
\rho_{b}\equiv \frac{\rho_b^0}{\eta\zeta a_m^3} \cdot \frac{e^{-a^2/a_m^2}}
{\left( 1+a/a_0 \right)^{1.8}},
\nonumber
\ee
\be
a\equiv \sqrt[]{x^2+\frac{y^2}{\eta^2} + \frac{z^2}{\zeta^2}}, \quad
R\equiv  \sqrt[]{x^2+y^2} \quad \textrm{and} \quad \vec{x}=(x,y,z).
\nonumber
\ee

Bissantz \& Gerhard \cite{bi02} added 4-armed spiral structure to this
density $\rho$, according to the model of Ortiz \& L\'epine \cite{or93},
and then used the model as initial model in their non-parametric de-projection. 
The positions of the spiral arms $r_i(\phi)$ ($i=1,\ldots,4$) 
in the model of Ortiz \& L\'epine \cite{or93} are given by
\be
r_i(\phi)=2.33\kpc \cdot e^{\left(\phi-\phibar-\phi_i \right) 
\cdot \tan(\chi)},
\nonumber\ee
where the angle $\phi_i=0,\pi/2,\pi, 3\pi/2$ determines the innermost position
angle of a spiral arm in Galacto-centric coordinates with respect to
the major axis of the bar, and $\chi=13.8\deg$ is the pitch angle of
the arms. 

In our parametric models the spirals exist between Galactocentric
radius of $3.5\kpc$, which is the approximate outer extend of the
bar/bulge (Bissantz \& Gerhard, 2002), and an outer radius 
of $10\kpc$. The spiral arms are
modelled by a Gaussian profile with full width at half maximum 
$\approx 300\pc$ (Ortiz \& L\'epine, 1993). 
We treat the spiral arms as enhancements of the disc density. 
The only free parameter that we fit for the spiral structure is 
the amplitude $d_{s}$ of the density modulations:
\be
\rho_d^{\rm including\ spiral}=\rho_d\cdot\prod_{i=1}^{4}
\left(1+d_s\cdot e^{-\ln(2)\cdot \Delta r_i^2/(0.5\cdot{\rm FWHM})^2} \right),
\nonumber\ee
where $\Delta r_i$ is the (approximate) distance to the nearest point
along spiral arm $i$. 
To keep the problem computationally tractable we use 
this rather simple model of the spiral structure in the
following. Note that Vall\'ee \cite{vallee2002}
suggests some improvements to the model of Ortiz \& L\'epine, however there
is good agreement between these models 
in the radial range $3\kpc\leq r\leq 6\kpc$ from the Galactic Center. This
is where most of the evidence results for spiral structure in the subsequently 
analysed region of the sky (cf. Sect. \ref{datagen}). Also
visual inspection of the residuals of all parametric models
showed no systematic deviations between the position of model spiral 
arm tangent points from those in the data.

We call this 4-armed model "model A" in the subsequent analysis. The other
parametric models under investigation 
are modifications of model "A". Model "B" has no
spiral arms at all, and in model "C" we omit the Sag-Car-arm and its counter-arm,
as suggested by Drimmel \cite{dr00}.
Finally, in our "intermediate" model "D" the amplitude $d_s$ of the Sag-Car 
arm and its counter-arm are a factor of $0.4$ smaller than the amplitude of the 
other pair of arms, as suggested by Drimmel \& Spergel \cite{dr01}. 

\subsection{The non-parametric model}
\label{nonpara}
As a non-parametric competitor to the above mentioned parametric models
we use the reference non-parametric luminosity density model of 
Bissantz \& Gerhard \cite{bi02}. 
It was estimated from the dust-corrected 
{\sl COBE/DIRBE} L-band map of Spergel et al.
\cite{sp95}, using a penalised maximum likelihood algorithm with 
penalty terms that encourage in the  model density 
eightfold-symmetry with respect to the three 
main planes of (approximate) symmetry of the bar, smoothness, and spiral 
structure  (similar to the 4-armed spiral
structure of the Ortiz \& L\'epine, 1993, model). 
The non-parametric model is defined on a (Galactocentric) grid of $60\!\times\! 
60\!\times\! 41$ grid points which extends $10\kpc$
along the $x$ and $y$ axis, and $3\kpc$ along the $z$-axis. Outside of this
box the model is continued by a parametric model of the L-band map.

%

Bissantz \& Gerhard (2002) derived non-parametric 
models for bar angles  $10\deg\leq\phibar\leq 44\deg$. They
found a preferred range of bar angles  
$20\deg\leq\phibar\leq 25\deg$. Their best model is for  
bar angle $\phibar\!=\!20\deg$, and is the "reference" model analysed 
subsequently.

\section{Application to models of the Milky Way}
\label{appl}
In this section we describe the application
of our statistical method to MW models A-D.
To this end we estimate $M^2$, which
is the distance between the ``true'' MW density $\rho$,
and the ``best-fitting'' parametric model $\rho_{\vartheta^{\ast}}$,
of each of the model (classes) A-D. Note that we use
the reference non-parametric model as an objective standard
of measurement. 
We approximately determine the statistic of $M^2$ by 
bootstrap replication of the distance $\hat M^2_{(i)}$ 
(in our case $i=1,\ldots,\approx 1000$) between ``best 
fitting'' models $\rho^{(i)}_{\hat\vartheta}$ of random data 
$\omega_{\rm obs}^{(i)}$ and the non-parametric reference 
model $\omega^{(np)}$ (cf. Sect. 3). 
The random data $\omega_{\rm obs}^{(i)}$ is generated 
from the {\sl COBE/DIRBE} data $\omega_{\rm obs}$ by drawing 
with replacement. 
We describe in this section the generation of the random sets of 
data (Sect. \ref{datagen}), the computation of $\hat M^2_{(i)}$ 
(Sect. \ref{ausfuehrung}), and finally, in Sect. \ref{eval}, we
explain how to evaluate the resulting statistic of $\hat{M}^2_{(i)}$ 
for the parametric models.

\subsection{Random data sets based on the {\sl COBE/DIRBE} data}
\label{datagen}
We construct random sets of data $\omega_{\rm obs}^{(i)}$
from a subset of the Spergel et al. \cite{sp95} {\sl COBE/DIRBE} data
which is generated in a first step as follows: 
We linearly interpolate the surface
brightness data (in magnitudes) on a grid ${\cal G}$
with $N=4800$ equidistant points,  covering $|l|\leq 40\deg$, 
$|b|\leq 10\deg$ from the observed data. 
We restrict the data to this area because
there the parametric continuation of the non-parametric model 
contributes only unimportantly to the projection of the model to 
the sky.

In the second step we estimate the local noise properties of
the {\sl COBE/DIRBE} data on the grid ${\cal G}$. To this
end we compute a map of the square difference between the 
projection of the non-parametric reference model to the sky and 
the {\sl COBE/DIRBE} data (both in magnitudes) at those
positions of the sky where this data is available. Then we smooth 
this map, employing a Savitzky-Golay filter (Press et al., 
1994, setting $n_R=n_L=M=2$ for their parameters). This
procedure yields a non-parametric estimate of the local variance 
$\hat\sigma^2(l,b)$ (cf. Sect. \ref{sec2}). 
Finally, we determine the local variance at 
the points of grid ${\cal G}$ by linear interpolation in this map.

In our final step we construct random data $\omega_{\rm obs}^{(i)}$ 
as follows. We randomly draw {\sl with replacement} $4800$ points 
$(l_j,b_j)$ out of the grid ${\cal G}$. The surface 
brightness observations $\omega_{\rm obs}(l_j,b_j)$ together with 
the estimates of the local variance $\hat\sigma^2(l_j,b_j)$, at the
drawn points $(l_i,b_j)$, then constitute a random
sample of data. Observe, that drawing with replacement implies that 
a point $(l_j,b_j)$ may occur more than once in 
this random sample.

\subsection{The distribution of $\hat M^2_{(i)}$}
\label{ausfuehrung}

We approximate the statistic of the distance between the best-fitting 
parametric model $\rho_{\hat\vartheta}$ for every parametric model 
A-D and the reference non-parametric model as follows. 
We compute $B\!=\! 1000$ bootstrap 
replications of $\hat{M}^2_{(i)}$, using the following procedure for
$1\leq i\leq 1000$:
\begin{enumerate}
\item
Generation of a random set of data $\omega_{\rm obs}^{(i)}$  from the
{\sl COBE/DIRBE} L-band data (cf. Sect. \ref{datagen}).
\item
Determination of the weighted least square estimator (WLSE) 
$\hat\vartheta^{(i)}$ (the best-fitting model) for this random data
set. For this we use a Marquardt-Levenberg-algorithm (Press et 
al., \cite{press}), which was used to minimize the distance 
$$\sum_{{\rm all}\, {\rm points} \,
(l,b)_j\, {\rm of} \, \omega_{\rm obs}^{(i)}} \frac{
\left(\omega_{\rm obs}^{(i)}\, ((l,b)_j)-\omega
_{\vartheta} ((l,b)_j)\right)^2}{\hat\sigma^2((l,b)_j)}.
$$ 
For the computation of $\hat\sigma^2$ see (ii) in Sect. 2.
The fitting is done in a two-step process:
\begin{description}
\item[1. Fitting of the disk parameters:]
In the first step we fit the disk parameters and the bulge
normalisation $b$, with the other bulge parameters fixed.
\item[2. Fitting of the bulge/bar parameters:]
In the second step we fix
the disk related parameters found in the first step (except for the
normalisation parameter $d$) 
and fit the bulge/bar parameters and $d$.
\end{description}
\item
Computation of the distance $\hat M^2_{(i)}$ 
between the reference non-parametric model
and the best-fitting parametric model $\hat\vartheta^{(i)}$:
\begin{eqnarray}
\hat M^2_{(i)} & = & ||\omega_{\hat\vartheta^{(i)}}
-\omega^{\rm (np)}||^2\nonumber\\
& = & \frac{1}{n\cdot m}\sum_{{\rm all}\, {\rm points}\atop
((l,b)_j)\,{\rm of}\, {\cal G}}
\left(\frac{\omega_{\hat\vartheta^{(i)}}((l,b)_j)
-\omega^{\rm (np)}((l,b)_j)}{\hat\sigma^2((l,b)_j)}\right)^2,\nonumber
\end{eqnarray}
where $n\cdot m=4800$. 
\end{enumerate}

\begin{figure}
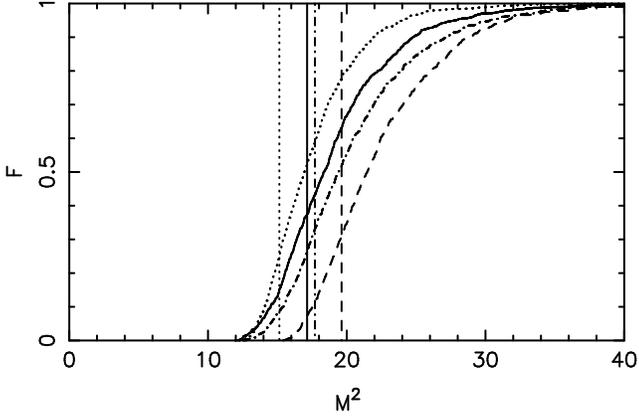

\getfig{plot1_all.eps}{6cm}
\caption{Cumulative distribution functions 
of $\hat{M}^2_{(i)}$
for models  A (full line), B (dashed), C (dot-dash-dot-dash) and
D (dotted). 
The vertical line indicates the distance $\hat{M}^2$ of the 
best-fitting parametric model $\rho_{\hat\vartheta}$ for the
original {\sl COBE/DIRBE} data from the non-parametric model.}
\label{paranonpara}
\end{figure}

From the $B\! =\! 1000$ bootstrap replications $\hat M_{(i)}^2$ we then
determine the empirical cumulative probability distribution function
of the random quantity $\hat{M}^2_{\ast}$,  shown in Fig. \ref{paranonpara},
for the parametric models A-D. Also indicated in Fig. \ref{paranonpara} are 
the estimated
distances $\hat{M}^2$ between the non-parametric model and the 
best-fitting parametric models $\rho_{\hat\vartheta}$ of the {\sl COBE/DIRBE} 
surface brightness, restricted to the grid ${\cal G}$, which is 
computed from the observed {\sl COBE/DIRBE} data. 

\subsection{$P$-value curves: Interpretation of the numerical results}
\label{eval}
\begin{figure}
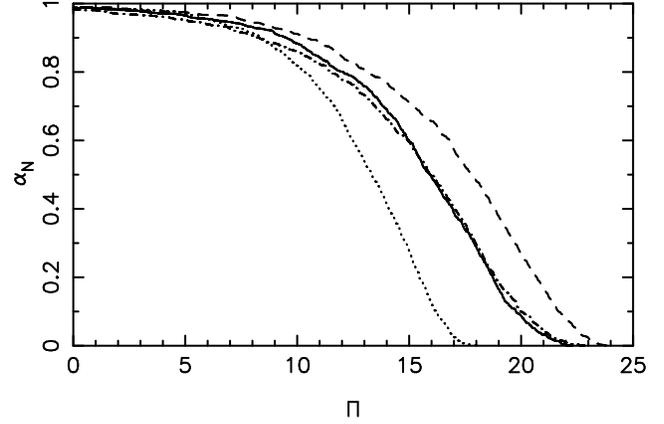

\getfig{plot3_all.eps}{6cm}
\caption{$P$-value curves $\alpha_N(\Pi)=F_B^{\ast}\left(
\hat{M}^2 - \Pi\right)$ for 
models A (full line), B (dashed), C (dot-dash-dot-dash), and D (dotted).}
\label{pvalue}
\end{figure}

We now proceed by using the bootstrap replications for the
parametric models A-D to answer two fundamental questions: 
\begin{enumerate}
\item
Should the non-parametric model be preferred over the parametric
models, i.e. does it improve the fit to the informative part (signal)
of the data?
\item
How do the parametric models compare to each other, can we decide
for a "best" among them, and - if yes - which one is it?
\end{enumerate}

We begin with some general remarks on our proposed method before we
use it to answer these questions. 
The main methodology we propose is the use of $P$-value curves
for the distance $M^2$
as graphical tools for illustrating the evidence for or against a model. 
To this end we plot the function $\alpha_N(\Pi)=F^{\ast}_B\left(\sqrt[]{N}
\left(\hat{M}^2-\Pi\right)\right)$ for $\Pi>0$, i.e. the value of $\alpha_N(\Pi)$
is given by the probability that the random quantity $X=\sqrt{N}\left(\hat M_{\ast}^2
-\hat M^2\right)$ is smaller than $\sqrt{N}\left(\hat M^2-\Pi\right)$. 
Note that this implies that for $\Pi$ increasing $\alpha_N(\Pi)$ decreases,
because we then evaluate the cumulative distribution function $F_B^{\ast}(x)$
for decreasing $x$, and in particular, if $\alpha_N(\Pi)$ is small, 
at the left tail of $F_B^{\ast}$. We present the $P$-value curves for
the parametric models A-D in Fig. \ref{pvalue}. 

The interpretation of the function $\alpha_N(\Pi)$ is as follows.
Assume the true distance between a parametric model  and the ``true'' density
$K\rho$ is $M^2=\Pi$. 
Now we reject the hypotheses $H:M^2>\Pi$ (vs. alternative $M^2\leq \Pi$)
whenever $\alpha_N(\Pi)\leq \alpha$ for a given level of significance $\alpha$.
Hence $1-\alpha_N(\Pi)$ can be regarded as the estimated evidence in favour of the
parametric model $U$ (up to a distance between parametric model and true density
$M^2\leq\Pi$). Finally the 
astrophysicist has to decide whether a value of $M^2=\Pi$ should be regarded as 
scientifically negligible or as deviation from the ``true'' density
$K\rho$ which is considered as significant by astrophysical reasons. 
For a more thorough introduction into $P$-value curves in the astrophysical 
context see [BM2], and for the statistical theory see
Munk (2002) Sect. 5 and the references given there.

With this interpretation in mind we now determine "distance margins" from the $P$-value 
curves in Fig. \ref{pvalue}. These margins indicate the most likely distance between 
the true density and the individual parametric models. We use those distances 
$\Pi$ where the error probability for the hypothesis "the distance between
parametric model and true density $K\rho$ is larger than $\Pi$" is  
$\approx\! 95\%$ and $\approx\! 5\%$ to define the upper and lower bound of the distance
margins (confidence interval), and find $6.5\! \leq\! M^2\! \leq\! 20.6$ for model A,
$7.7\! \leq\! M^2\! \leq\! 22.5$ for model B, 
$5.1\! \leq\! M^2\! \leq\! 20.9$ for model C, and
and $6.3\! \leq\! M^2\! \leq\! 16.7$ for model D. 

Observe that our distance margins are much more meaningful than, say, computing the
``classical'' $\chi^2$-distance  
between parametric model and data.  This is in particular
because the distance margins not only give the absolute distance between the parametric
model and the non-parametric one, but also error bounds for the distance.
Furthermore, our method allows and adapts for the commonly poorly known and heteroscedastic
distribution of observational noise - in contrast to classical $\chi^2-$methods.

It is important to comment
on the validity of our approach of using the non-parametric 
model as an ``objective standard of measurement'' in the context of the 
quality of the non-parametric model. To this end we consider two extreme
cases. First, assume that the non-parametric model would fit the data rather 
bad relative to the estimated local standard deviation $\hat \sigma(l,b)$. 
In consequence the cumulative distribution function of
$\hat M^2_{(i)}$ tends sharply to $1$. This implies a $P$-value curve which
goes to zero very quickly,  and we have to conclude that the 
distance between parametric model and ``true'' density $K\rho$ is very small,
and hence there is no reason to prefer non-parametric over parametric modelling
of the data.
On the other hand, now assume that the non-parametric model is a (nearly) perfect
fit to the data relative to the local standard deviation. Then the
cumulative distribution function of $\hat M^2_{(i)}$, i.e. of the estimated distance 
between parametric model and ``true'' density, tends very slowly to $1$,
implying a $P$-value curve which decreases to zero rather slowly. In this
case we cannot make any statement about a small distance of the parametric
model from the ``true'' density.
In conclusion our new method even remains valid for ``bad'' non-parametric models
serving as ``objective standards of measurement''. 

We now use the distance margins to answer our first question - whether the
non-parametric model should be prefered over the parametric models. To this end
we have to decide whether we consider the determined distance margin between a parametric 
model and the true density (here represented by the non-parametric model which is
used as an objective standard of measurement) as astrophysically significant. 
For this we compare the distance margins with crude estimates for the distance between
the non-parametric model and the {\sl COBE/DIRBE} data on grid ${\cal G}$, 
and for the error dispersion
in the {\sl COBE/DIRBE} L-band data, as given by Spergel et al. \cite{sp95}. 
If these latter values are smaller than the typical distance margins we conclude 
that the latter distances are astrophysically significant and the non-parametric model 
should be prefered. 

\begin{figure*}
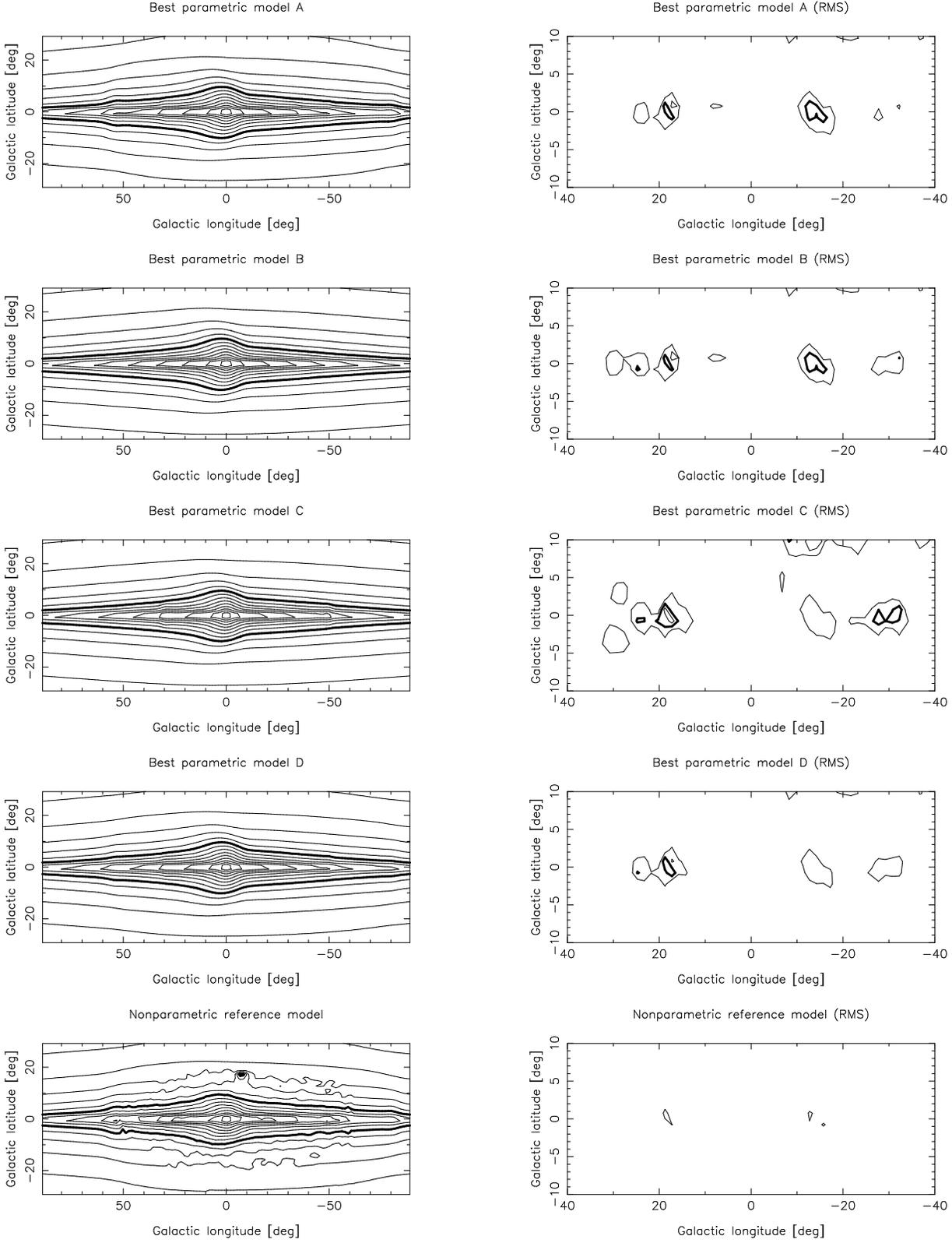

\begin{centering}
\get2fig{newfig4.ps}{16.15cm}
\end{centering}
\caption{Square (rms) difference between the best-fitting parametric 
models $\rho_{\hat\vartheta}$  and the {\sl COBE/DIRBE} 
L-band data, and the same difference for the reference non-parametric model. 
Contour spacing is $0.05 {\rm mag}^2$ and 
the bold contour is at $0.1 {\rm mag}$ for 
for the rms plots (left hand side) and $0.5 {\rm mag}$, $0 {\rm mag}^2$
for the model plots (right hand side). Contour lines in the rms plots
have been smoothed by averaging over a point and its four nearest 
neighbours. Different areas of the sky are shown for the rms plots 
(the region of sky to which the parametric models were fitted) and
the model plots (full sky area covered by our {\sl COBE/DIRBE} map).
Note the different appearance of spiral arm tangent points for
the different models, and the corresponding residuals in the rms
maps.}
\label{rmsplot}
\end{figure*}

We begin with the computation of the distance between the non-parametric model 
and the {\sl COBE/DIRBE} data on the grid ${\cal G}$, normalized by the estimated 
variance of the {\sl COBE/DIRBE} L-band data, which is $\approx 0.076^m$ 
(Spergel et al., 1995, cf. Bissantz \& Gerhard, 2002). This normalisation
enables us to compare the above determined distances between the non-parametric model
and the parametric models, observing that the quantity $\hat{M}^2$ is similar in 
construction to a rms difference between the models (on the sky), 
``weighted'' by the local variance of the data $\hat\sigma^2(l,b)$. We find $0.62$ 
for our crude estimate of the distance between the reference non-parametric model 
and the observed data. This, and in particular the normalized dispersion of the data 
(which by our definition amounts to $1$), is significantly less than the 
distance margins indicate for the parametric models. Thus we conclude
that the non-parametric model improves significantly over the parametric models
and should be prefered. 

This conclusion is supported by Fig. \ref{rmsplot} where the squared (rms)
residuals between the reference non-parametric model and the {\sl COBE/DIRBE}
L-band map, and the same for the best parametric models $\rho_{\hat\vartheta}$ 
(the best of which is of type D, cf. Fig. \ref{paranonpara}), 
are compared. Obviously the
non-parametric model fits the data very well, in particular 
in the central region $|l|\leq 50\deg, |b|\leq 20\deg$. The 
residuals of the parametric model, however, show systematic
deviations from the data.  Note, however, that the $P$-value curves provide
us with substantially more information than these maps of model residuals,
because they guard us against overfitting in a quantitative way.


Having decided that the  
distance margins are of astrophysical significance we proceed to the second question,
which is to compare the parametric models among each others. This will be done for
illustrational purpose mainly, because we have already seen that a non-parametric
model should be prefered. To this end we 
use the right tail of the $P$-value curves, where $\alpha_N(\Pi_r)\!\approx\! 0.05$, 
i.e. we determine the distance $\Pi_r$ for which the hypothesis "the distance between 
the parametric model and the true density $K\rho$ is larger than $\Pi_r$" has an 
error probability $\approx\! 95\%$. Inspection of Fig. \ref{pvalue} shows that the 
intermediate model D outperforms 
the other parametric models because $\Pi_r$ is smallest for this model, indicating
that it is the best among models A-D. The 2- and  4-armed models A and C are approximately
similar in quality, and the model without spiral arms (B) is clearly the worst. 
We remark that the differences between the distances $\Pi_r$ for the models (as given
by the upper bound of the distance margins, cf. above) are significantly larger
than the normalized distance between the non-parametric model and the data, and
in particular than the normalized dispersion in the data. We conclude that 
the differences between the models are significant, and that the intermediate model D is
best, i.e. we can exclude a large distance of the parametric model from the
true MW density with a higher level of confidence for model D than for models A-C.

\begin{figure}
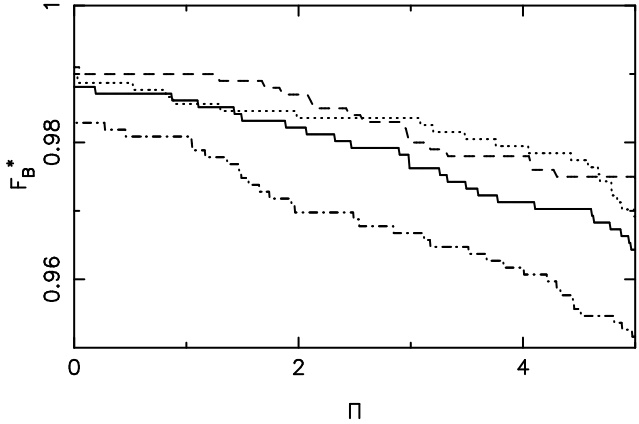

\getfig{plot3_scaled.eps}{6cm}
\caption{$P$-value curves 
$\alpha_N(\Pi)=F_B^{\ast}\left(
\hat{M}^2 - \Pi\right)$ for 
models A (full line), B (dashed), C (dot-dash-dot-dash), and D (dotted),
respectively, where $\Pi\approx 0$.}
\label{pvaluecloseup}
\end{figure}

Finally, we comment on the $P$-value curve analysis in
the context of "classical" statistical tests, which start off from the hypothesis
$\Pi=0$. Fig. \ref{pvaluecloseup} presents the left tail of our $P$-value curves, 
where $\Pi\approx 0$, showing that some of the $P$-value curves intersect. 
This has an interesting consequence:  a "classical" test based on the 
hypothesis that the model holds ($\Pi=0$)  leads to a different answer (here: model
C performs best) than our proposed method, which is based on the more realistic
hypothesis, that there is a nonzero distance $\Pi>0$ between the parametric model and
the true MW density $K\rho$.
Observe that, because we have found in our precedent analysis that the observed
distances between the parametric models and the true MW density are significant, 
a classical test would indeed be inappropriate in our application. 

To close this section we summarize the answer to our questions raised at the 
beginning of this section:
\begin{enumerate}
\item
The non-parametric model should be prefered over the parametric models because the 
improvement in fit is astrophysically relevant and 
with high confidence due to systematic features. 
\item
A parametric model with 4-armed spiral structure, however, with reduced amplitude
of the Sagittarius-Carina arm and its counter-arm as suggested by Drimmel \& Spergel
\cite{dr01}, significantly outperforms the other parametric models. The worst
performance is shown by a model without spiral arms at all as to be expected. 
\end{enumerate}

\section{Discussion}

We have suggested a resampling algorithm to assess the necessity of non-parametric
instead of parametric modelling in astrophysical (inverse) regression problems.
By means of this we are in the position 
to decide whether the deviations between the parametric model and
the data
are systematic or due to noise. Furthermore our method can be used to select the
best among several competing parametric models. 
Our approach is based on the idea to investigate the statistical behaviour of 
the estimated distance between the true model $\rho$ and the artifical model
$U$ under all "possible worlds" $M^2=\Pi$, and not only when $M^2=0$ (i.e. 
$\rho$ and $U$ coincide), as
classical goodness of fit tests do. Moreover, we compare all
parametric models by relating them to a non-parametric "super-model" which 
can be validated itself by our method.

To illustrate our method we have applied it
to the problem of recovering the near-infrared
luminosity density distribution of the Milky Way from a dust-corrected {\sl
COBE/DIRBE} L-band map (Spergel et al., 1995). In this paper we have 
focused on the morphology of the spiral arms, comparing parametric models
which have zero, 2 or 4  spiral arms, and also an "intermediate" 4-armed model,
in which the Sagittarius-Carina arm and its counter-arm are considerably less 
strong than the other pair of arms. These parametric models have been compared
with a non-parametric model of Bissantz \& Gerhard 
\cite{bi02}. From our statistical analysis we conclude that the
non-parametric model is significantly better than the parametric models
and hence should be prefered. This is due to systematic departures between 
data and parametric models, which are in particular to inflexible to 
reproduce certain deviations from a double-exponential disk and smooth spiral 
arms.
Furthermore, we have found that the "intermediate"
parametric model outperforms the other parametric models by a significant
amount. Thus, from the analysis of the dust-corrected {\sl COBE/DIRBE} 
L-band map, a parametric model of the Milky Way with 4-arms, but
the Sagittarius-Carina arm (and in our model also its counter-arm)
of reduced amplitude - similar to the suggestion 
of Drimmel \& Spergel \cite{dr01} - is to be prefered over models with  
2-armed or 4-armed structure, and in particular over a model without
spiral structure, which performed worst in our analysis.

Before making our final conclusions we want to point out some
difficulties of the {\sl COBE/DIRBE} NIR data 
with respect to spiral arm analysis as pointed out by a referee. Firstly, 
dust extinction, which is biased towards the spiral arms, makes those less 
evident in NIR data (cf. Drimmel \& Spergel, 2001). Our analysed L-band
map was dust-corrected by Spergel et al. \cite{sp95},
but this extinction-bias obviously also results in their dust-correction 
being more difficult to perform. Secondly,
the large angular size of the nearby Sagittarius-Carina
arm makes it more difficult to observe in NIR maps, 
because combined with the small instrument beam a larger amount of emission
will be lost below the sensitivity threshold of the instrument than for
the other arms. Finally, since we had to restrict our parametric fitting
to a range of longitudes $|l|\leq 40\deg$ of the data due to computational
reasons (range of the non-parametric model), we did not include the tangent
point regions of the Sagittarius-Carina arm, which substantially weakens
any conclusion regarding a differing amplitude of this arm. 
The latter two points obviously weaken the statistical evidence for 
the intermediate model compared to the 4-armed model by some amount.

This result is consistent with Drimmel \& Spergel \cite{dr01}, and
also with the SPH models of the inner Milky Way gas dynamics of 
Bissantz \& Gerhard \cite{bi02b}, who found that 4-armed spiral structure
is needed in the gravitational potential caused by the stellar distribution. 
They analysed the gas flow in a potential ("mix"), which is quite similar 
to our intermediate model. This gas model was found to be slightly worse 
than gas flow models in their standard 4-armed potential, but not nearly 
as bad as the gas flow in 2-armed potentials, and is clearly still acceptable. 

\section{Acknowledgments}
We are grateful to O.Gerhard and H.Dette for helpful comments, and thank
the anonymous referees for thoughtful suggestions which lead to a clearer 
presentation of our result. The source code of our method can be obtained
from the authors on request. Please send an e-mail to bissantz@math.uni-goettingen.de

\end{document}